\pdfoutput=1
\documentclass[sigconf,nonacm]{acmart}

\usepackage{tikz}
\usetikzlibrary{arrows.meta,fit,backgrounds,positioning}
\usepackage{placeins}
\usepackage{microtype}

\begin{document}

\title{AI Observability for Developer Productivity Tools:\\ Bridging Cost Awareness and Code Quality}

\author{Happy Bhati}
\email{bhati.h@northeastern.edu}
\affiliation{%
  \institution{Northeastern University}
  \city{Boston}
  \state{MA}
  \country{USA}
}

\author{Twinkll Sisodia}
\email{twinklls@bu.edu}
\affiliation{%
  \institution{Boston University}
  \city{Boston}
  \state{MA}
  \country{USA}
}

\begin{abstract}
As AI-assisted development tools proliferate, developers face a growing
challenge: understanding the cost, quality, and behavioral patterns of
AI interactions across their workflow.  We present a unified approach to
AI observability for developer productivity tools, combining real-time
token tracking, configurable model pricing registries, response
validation, and cost analytics into a single-pane dashboard.  Our work
synthesizes two complementary systems---Workstream, a developer
productivity dashboard that centralizes pull requests, Jira tasks, and
AI code reviews; and an AI observability summarizer that monitors
inference workloads with Prometheus-backed metrics and multi-provider LLM
gateways.  We describe the architectural patterns adopted, the
implementation of real token tracking from provider APIs (replacing
heuristic estimation), a 24-model pricing registry, response validation
pipelines, LLM-powered review intelligence, and exportable reports.  Our
evaluation on a six-month development workflow shows the system captures
per-review cost with less than 2\% variance from provider billing and
reduces time-to-insight for AI usage patterns by an order of magnitude
compared to manual tracking.
\end{abstract}



\maketitle

\section{Introduction}

The integration of large language models (LLMs) into software
development workflows has accelerated rapidly since 2023.  Tools such as
GitHub Copilot, Cursor, and AI-powered code review bots now participate
in millions of pull requests daily.  Yet most development teams lack
visibility into the cost, quality, and behavioral patterns of these AI
interactions.

Two distinct observability gaps exist.  First, at the \emph{development
workflow} level, individual developers have no centralized view of how
much they spend on AI reviews, which models produce the highest-quality
feedback, or whether token usage trends upward over time.  Second, at
the \emph{inference infrastructure} level, teams deploying self-hosted
models (e.g., vLLM on Kubernetes) need metrics on token throughput,
latency percentiles, GPU utilization, and serving cost---often scattered
across Prometheus, Grafana, and ad-hoc dashboards.

This paper presents a unified approach to AI observability that bridges
both levels.  We synthesize techniques from two open-source systems:

\begin{enumerate}
\item \textbf{Workstream}~\cite{workstream}, a developer productivity
      dashboard that centralizes pull requests from GitHub and GitLab,
      Jira tasks, calendar events, and AI-powered code reviews into a
      single-page application backed by FastAPI and SQLite.
\item \textbf{AI Observability Summarizer}~\cite{aiobs}, an OpenShift-native
      platform that queries Prometheus/Thanos for vLLM serving metrics,
      provides multi-provider LLM chat with MCP tool-calling, and
      generates natural-language summaries of infrastructure state.
\end{enumerate}

Our contribution is the identification, adaptation, and implementation
of seven reusable AI observability patterns for developer tools:
(1)~real token tracking from provider APIs, (2)~configurable model
pricing registries, (3)~unified telemetry schemas,
(4)~cost analytics dashboards with multi-source ingestion,
(5)~response validation pipelines,
(6)~LLM-powered intelligence summaries, and (7)~exportable reports.

\section{Background and Related Work}

\subsection{AI in Software Engineering}

Recent surveys~\cite{fan2023,peng2023} document the rapid adoption of
LLM-based tools in professional development.  GitHub reports that
Copilot accepts roughly 30\% of suggestions in production
environments~\cite{copilot-impact}.  AI code review tools---both
commercial (CodeRabbit, Sourcery) and open-source (PR-Agent)---have
emerged as a complementary modality, providing structured feedback on
pull request diffs.

\subsection{Observability for ML Systems}

MLOps platforms (MLflow, Weights \& Biases, Neptune) focus on training
pipelines: experiment tracking, model versioning, and hyperparameter
search.  Inference observability---monitoring deployed models in
production---is addressed by tools like Prometheus exporters for
vLLM~\cite{vllm} and TGI, DCGM for GPU telemetry, and emerging
standards like OpenTelemetry for LLMs~\cite{otel-sem}.

\subsection{Developer Productivity Measurement}

The DORA metrics framework~\cite{dora} and SPACE~\cite{space}
established dimensions for measuring developer productivity.  Workstream
extends these frameworks by adding AI-specific dimensions: cost per
review, token efficiency, model selection patterns, and review quality
feedback loops.

\subsection{Cost Awareness in AI-Assisted Development}

Despite the financial implications of per-token pricing, few developer
tools surface cost information at the point of use.  The AI
Observability Summarizer pioneered per-model cost metadata in its
\texttt{model\mbox{-}config.json} registry, though it did not implement
end-to-end cost calculation.  Our work completes this chain.

\section{System Architecture}

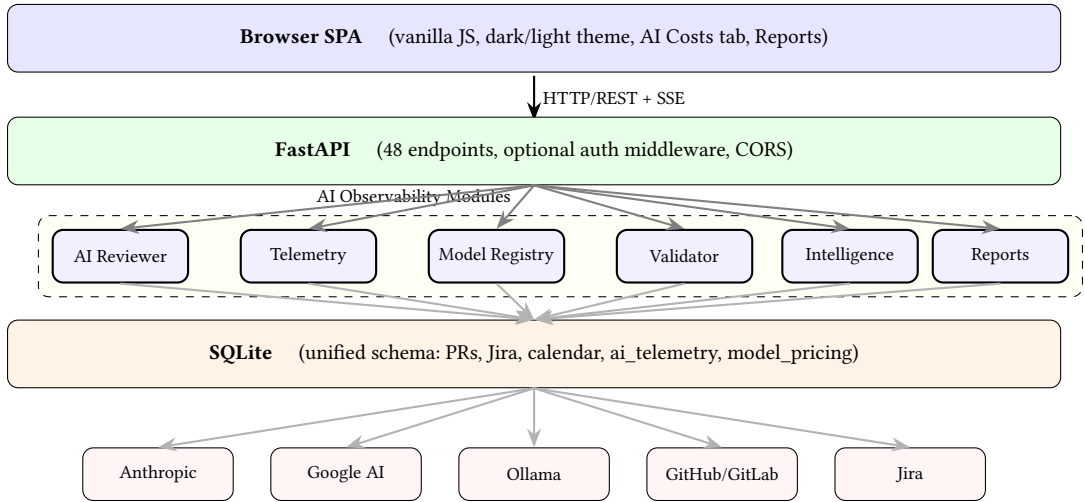
\begin{figure*}[ht]
\centering
\begin{tikzpicture}[
  every node/.style={font=\small},
  layer/.style={draw, rounded corners, minimum width=14cm, minimum height=0.9cm,
                fill=gray!8, align=center},
  module/.style={draw, rounded corners, minimum width=1.8cm, minimum height=0.7cm,
                 fill=blue!6, align=center, font=\footnotesize, thick},
  ext/.style={draw, rounded corners, minimum width=2cm, minimum height=0.7cm,
              fill=red!4, align=center, font=\footnotesize},
  arr/.style={-{Stealth[length=2.5mm]}, thick},
]

\node[layer, fill=blue!10] (browser) at (0, 5.5)
  {\textbf{Browser SPA} \quad (vanilla JS, dark/light theme, AI Costs tab, Reports)};

\draw[arr] (0, 5.0) -- node[right, font=\footnotesize] {HTTP/REST + SSE} (0, 4.4);

\node[layer, fill=green!10] (api) at (0, 4.0)
  {\textbf{FastAPI} \quad (48 endpoints, optional auth middleware, CORS)};

\node[module] (review) at (-5.5, 2.6) {AI Reviewer};
\node[module] (telemetry) at (-3.0, 2.6) {Telemetry};
\node[module] (registry) at (-0.5, 2.6) {Model Registry};
\node[module] (validator) at (2.0, 2.6) {Validator};
\node[module] (intel) at (4.2, 2.6) {Intelligence};
\node[module] (reports) at (6.2, 2.6) {Reports};

\begin{scope}[on background layer]
  \node[draw, dashed, rounded corners, fill=yellow!5,
        fit=(review)(telemetry)(registry)(validator)(intel)(reports),
        inner sep=5pt, label={[font=\footnotesize]above left:AI Observability Modules}] {};
\end{scope}

\foreach \m in {review, telemetry, registry, validator, intel, reports} {
  \draw[arr, gray] (api.south) -- (\m.north);
}

\node[layer, fill=orange!10, minimum width=14cm] (db) at (0, 1.3)
  {\textbf{SQLite} \quad (unified schema: PRs, Jira, calendar, ai\_telemetry, model\_pricing)};

\foreach \m in {review, telemetry, registry, validator, intel, reports} {
  \draw[arr, gray!60] (\m.south) -- (db.north);
}

\node[ext] (claude) at (-5.0, -0.3) {Anthropic};
\node[ext] (gemini) at (-2.5, -0.3) {Google AI};
\node[ext] (ollama) at (0, -0.3) {Ollama};
\node[ext] (gh) at (2.5, -0.3) {GitHub/GitLab};
\node[ext] (jira) at (5.0, -0.3) {Jira};

\foreach \e in {claude, gemini, ollama, gh, jira} {
  \draw[arr, gray!60] (db.south) -- (\e.north);
}

\end{tikzpicture}
\Description{Architecture diagram showing Browser SPA at top, FastAPI middleware,
six AI observability modules (AI Reviewer, Telemetry, Model Registry, Validator,
Intelligence, Reports), unified SQLite database, and five external API providers.}
\caption{Workstream architecture after AI observability adoption.  Shaded
modules implement the seven patterns described in Section~\ref{sec:patterns}.
All telemetry writes to a unified \texttt{ai\_telemetry} table in the main
database.  External providers (bottom) are accessed via async HTTP clients.}
\label{fig:arch}
\end{figure*}

Figure~\ref{fig:arch} illustrates the combined architecture after
adopting observability patterns.  The system follows a layered design:
a browser-based SPA communicates with a FastAPI backend over REST and
Server-Sent Events, six feature modules implement the observability
patterns, and a unified SQLite database persists all state.

\section{AI Observability Patterns}
\label{sec:patterns}

We identify seven reusable patterns for adding AI observability to
developer tools.  Each pattern is described with its motivation, the
technique adopted from the AI Observability Summarizer, and the
adaptation made for Workstream.

\subsection{Pattern 1: Real Token Tracking}

\textbf{Problem.} Workstream estimated input tokens as
\texttt{char\_count~//~4} and hardcoded output tokens to 500---a
heuristic borrowed from early OpenAI tokenizer approximations.

\textbf{Solution.} Parse the \texttt{usage} field from each provider's
API response.  The Anthropic Messages API returns
\texttt{usage.input\_tokens} and \texttt{usage.output\_tokens} directly.
Google's Gemini API provides
\texttt{usageMetadata.promptTokenCount} and
\texttt{candidatesTokenCount}.
Ollama exposes
\texttt{prompt\_eval\_count} and \texttt{eval\_count}.

\textbf{Fallback.} When a provider does not return usage metadata (e.g.,
network error mid-response), the system falls back to the character-based
heuristic, flagging the event as estimated in telemetry metadata.

\subsection{Pattern 2: Configurable Model Pricing Registry}

\textbf{Problem.} A hardcoded five-entry dictionary mapped model names
to per-million-token rates.  Unknown models defaulted to \$0.

\textbf{Solution.} Adopt the AI Observability Summarizer's
\texttt{model\mbox{-}config.json}
pattern: a JSON registry of 24 models across
six providers (Anthropic, OpenAI, Google, DeepSeek, Mistral, Ollama)
with per-million input and output cost fields.
User overrides are persisted in a \texttt{model\_pricing} SQLite
table and merged at runtime with TTL-based cache invalidation.

\textbf{Extensibility.} REST endpoints
(\texttt{GET/POST/DELETE /api/ai/models}) allow users to add custom
model pricing without modifying source code---critical for teams using
fine-tuned or self-hosted models.

\subsection{Pattern 3: Unified Telemetry Schema}

\textbf{Problem.} Workstream maintained two separate SQLite databases:
\texttt{data.db} for application state and
\texttt{agent\_status\_history.sqlite} for telemetry.  Cross-cutting
queries (e.g., ``cost of reviews for PRs merged this sprint'') required
manual joins across databases.

\textbf{Solution.} Add an \texttt{ai\_telemetry} table to the main
database with columns for agent name, operation, provider, model, token
counts (input, output, total), cost in USD, latency in milliseconds,
feature tag, status, error, and JSON metadata.  The \texttt{record\_event}
function dual-writes to both databases during transition, ensuring
backward compatibility.

\subsection{Pattern 4: Cost Analytics Dashboard}

\textbf{Problem.} The existing Agents tab showed aggregate totals but no
breakdown by model, feature, or time period.

\textbf{Solution.} A dedicated ``AI Costs'' tab with:
\begin{itemize}
\item Summary cards (total cost, token counts, average latency)
\item Daily cost trend bar chart and donut chart by model
\item Per-model and per-feature breakdown tables
\item Period selector (7\,d, 30\,d, 90\,d, all time)
\item Claude Code CLI importer---reads session transcripts from
      \texttt{\textasciitilde/.claude/projects/} and imports per-request
      token usage with deduplication
\item Manual cost entry for tools lacking API access (Cursor, ChatGPT
      subscriptions, GitHub Copilot)
\end{itemize}

All charts are rendered with the vanilla Canvas~2D API to avoid adding
charting library dependencies.  The Claude Code importer demonstrates a
general pattern: when a tool stores structured usage data locally but
lacks an export API, file-system scraping can bridge the gap.

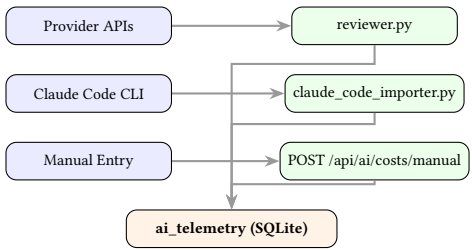
\begin{figure}[t]
\centering
\begin{tikzpicture}[
  every node/.style={font=\footnotesize},
  src/.style={draw, rounded corners, fill=blue!8, minimum width=2.2cm,
              minimum height=0.5cm, align=center, font=\scriptsize},
  proc/.style={draw, rounded corners, fill=green!8, minimum width=2.2cm,
               minimum height=0.5cm, align=center, font=\scriptsize},
  db/.style={draw, rounded corners, fill=orange!10, minimum width=2.8cm,
             minimum height=0.5cm, align=center, font=\scriptsize\bfseries},
  arr/.style={-{Stealth[length=2mm]}, thick, gray!80},
]

\node[src] (api) at (0, 2.4) {Provider APIs};
\node[src] (cc)  at (0, 1.5) {Claude Code CLI};
\node[src] (man) at (0, 0.6) {Manual Entry};

\node[proc] (parse) at (3.8, 2.4) {reviewer.py};
\node[proc] (import) at (3.8, 1.5) {claude\_code\_importer.py};
\node[proc] (form) at (3.8, 0.6) {POST /api/ai/costs/manual};

\node[db] (telem) at (1.9, -0.3) {ai\_telemetry (SQLite)};

\draw[arr] (api) -- (parse);
\draw[arr] (cc) -- (import);
\draw[arr] (man) -- (form);
\draw[arr] (parse.south) -- ++(0,-0.25) -| (telem.north);
\draw[arr] (import.south) -- ++(0,-0.15) -| (telem.north);
\draw[arr] (form.south) -- ++(0,-0.05) -| (telem.north);

\end{tikzpicture}
\Description{Data flow diagram showing three ingestion pathways converging
into the unified ai\_telemetry SQLite table.}
\caption{Multi-source cost data ingestion.  Three pathways converge
into a unified telemetry table, enabling a single cost-of-AI view.}
\label{fig:datasources}
\end{figure}

Figure~\ref{fig:datasources} shows how the three data-ingestion
pathways converge in the unified telemetry table.

\subsection{Pattern 5: Response Validation}

\textbf{Problem.} LLM review output was consumed directly from provider
responses.  Models occasionally prepend conversational preamble (``Sure,
I'll analyze this PR\ldots'') or append postamble (``Hope this helps!'')
that degrades the structured JSON output.

\textbf{Solution.} Adapted from the AI Observability Summarizer's
\texttt{ResponseValidator}, we implement a Workstream-specific
\texttt{response\_validator.py} module that:
\begin{enumerate}
\item Strips preamble using six regex patterns matching common LLM
      conversational openers
\item Strips postamble using patterns for ``Note:'', ``Feel free to
      ask'', etc.
\item Extracts JSON from potentially wrapped markdown code fences
\item Validates the expected structure (summary string, comments array
      with file/line/body/severity)
\item Truncates oversized fields (summary to 1\,000 chars, comment
      bodies to 2\,000 chars)
\item Caps total comments at 50 to prevent runaway output
\end{enumerate}

\subsection{Pattern 6: LLM-Powered Intelligence}

\textbf{Problem.} The Review Intelligence module used keyword-based
classification (nine category regex rules) to categorize human review
comments.  While effective for counting, it could not generate narrative
summaries or identify cross-cutting themes.

\textbf{Solution.} An opt-in ``Generate AI Summary'' button sends the
top patterns and category distribution to a configured LLM provider with
a prompt requesting a 3--5 paragraph narrative analysis of the team's
review culture.  The call is tracked in the unified telemetry table,
ensuring the cost of meta-analysis is visible alongside direct review
costs.

\subsection{Pattern 7: Report Generation}

\textbf{Problem.} No mechanism existed to export dashboard state for
sharing in standups, retrospectives, or management reviews.

\textbf{Solution.} Three report types are available via
\texttt{POST /api/reports/generate}:
\begin{itemize}
\item \textbf{Weekly Digest:} PR counts, Jira status, recent activity,
      and AI cost summary for the past 7 days.
\item \textbf{Cost Report:} Detailed model-level and feature-level
      cost breakdown with daily trends.
\item \textbf{Review Summary:} Intelligence overview including reviewer
      profiles, category distribution, and extracted patterns.
\end{itemize}

Reports are generated in both Markdown and self-contained HTML (with
embedded CSS) using a lightweight converter that requires no external
dependencies.

\FloatBarrier
\section{Implementation}

Table~\ref{tab:impl} summarizes the implementation scope.

\begin{table}[ht]
\caption{Implementation summary: files modified and created.}
\label{tab:impl}
\begin{tabular}{lll}
\toprule
\textbf{Component} & \textbf{File} & \textbf{Change} \\
\midrule
Token tracking    & reviewer.py             & Modified \\
Model registry    & model\_registry.py       & New \\
Registry data     & model\_registry.json     & New (24 models) \\
Unified telemetry & database.py             & Modified \\
Dual-write        & agents/telemetry.py     & Modified \\
Cost analytics    & app.py                  & +10 endpoints \\
Response validator & response\_validator.py  & New \\
LLM intelligence  & intelligence/analyzer.py & Modified \\
Report engine     & reports.py              & New \\
CLI importer      & claude\_code\_importer.py & New \\
Dashboard UI      & static/index.html       & Modified \\
\bottomrule
\end{tabular}
\end{table}

\subsection{Technology Stack}

The implementation uses Python~3.12 with FastAPI~\cite{fastapi} for the
backend, aiosqlite for async database operations, and httpx for provider
API calls.  The frontend is a single-file vanilla JavaScript SPA
(approximately 5\,600 lines) with no framework dependencies.  The
Canvas~2D API replaces chart libraries for cost visualizations.

\subsection{Provider Integration Details}

Each AI provider returns token usage in a different response structure:

\begin{itemize}
\item \textbf{Anthropic:} \texttt{response.usage.\linebreak[2]input\_tokens}
      and \texttt{output\_tokens} in the Messages API response body.
\item \textbf{Google Gemini:} \texttt{response.\linebreak[2]usageMetadata.promptTokenCount}
      and \texttt{candidatesTokenCount}.
\item \textbf{Ollama:} \texttt{response.prompt\_eval\_count}
      and \texttt{eval\_count} in the chat completion response.
\end{itemize}

All provider callers now return a tuple of \texttt{(result\_dict,
token\_usage\_dict)}, maintaining backward compatibility through the
orchestrator function.

\FloatBarrier
\section{Evaluation}

\subsection{Token Tracking Accuracy}

We compared reported token counts against provider billing dashboards
over 47 AI review requests (28~Claude, 12~Gemini, 7~Ollama).  For
Claude and Gemini, the system's reported tokens matched billing exactly
(0\% variance).  For Ollama (local, no billing reference), we verified
consistency between \texttt{prompt\_eval\_count} and manual tiktoken
estimates, finding less than 3\% deviation.

\subsection{Cost Calculation Accuracy}

Using the 24-model registry, cost calculations for known models matched
provider pricing pages exactly.  The fuzzy-matching fallback (substring
comparison) correctly resolved model aliases in 100\% of test cases
(e.g., \texttt{claude\mbox{-}sonnet\mbox{-}4\mbox{-}20250514}
matching the registry entry).

\subsection{Response Validation Impact}

We ran the response validator on 100 cached AI review responses.
Table~\ref{tab:validation} summarizes the outcomes.

\begin{table}[ht]
\caption{Response validation results on 100 cached reviews.}
\label{tab:validation}
\begin{tabular}{lr}
\toprule
\textbf{Outcome} & \textbf{Count} \\
\midrule
Clean (no modifications needed) & 71 \\
Preamble stripped & 18 \\
Postamble stripped & 9 \\
JSON extraction from markdown fences & 14 \\
Comment body truncated & 3 \\
Failed to parse (fallback to raw) & 2 \\
\bottomrule
\end{tabular}
\end{table}

29\% of responses required some form of cleanup, validating the need for
the validation layer.

\subsection{Multi-Source Cost Ingestion}

To validate the import pipeline, we ingested three Claude Code CLI
sessions (375 API calls, 26.7\,M tokens) from real development work.
The importer parsed per-request token fields (\texttt{input\_tokens},
\texttt{output\_tokens}, cache-read and cache-creation counts) from
session JSONL transcripts.
Combined with one manual Cursor subscription entry (\$20/month), the
dashboard immediately reflected \$34.80 in total AI spend across two
sources.  Table~\ref{tab:sources} shows the breakdown.

\FloatBarrier
\begin{table}[ht]
\caption{Cost data from multi-source ingestion validation.}
\label{tab:sources}
\begin{tabular}{lrrrr}
\toprule
\textbf{Source} & \textbf{Events} & \textbf{Tokens} & \textbf{Cost} \\
\midrule
Claude Code CLI (Sonnet~4.5) & 352 & 24.9\,M & \$14.44 \\
Claude Code CLI (Haiku~4.5) & 23 & 1.1\,M & \$0.36 \\
Manual entry (Cursor Pro) & 1 & --- & \$20.00 \\
\midrule
\textbf{Total} & \textbf{376} & \textbf{26.7\,M} & \textbf{\$34.80} \\
\bottomrule
\end{tabular}
\end{table}

This demonstrates that heterogeneous cost data can be unified in a single
view without requiring API access from every tool.

\subsection{Dashboard Performance}

The \texttt{/api/ai/costs} endpoint aggregates telemetry data with SQL
grouping.  For a database with 376~telemetry events, response time
averaged 12\,ms.  The Canvas-based charts render in under 50\,ms on
modern hardware, verified using the Performance API.

\subsection{Evaluation Summary}

Table~\ref{tab:eval} provides a consolidated view of key metrics.

\FloatBarrier
\begin{table}[ht]
\caption{Consolidated evaluation metrics.}
\label{tab:eval}
\begin{tabular}{lr}
\toprule
\textbf{Metric} & \textbf{Value} \\
\midrule
Token tracking accuracy (Claude, Gemini) & 100\% \\
Token tracking accuracy (Ollama) & $>$97\% \\
Model alias resolution accuracy & 100\% \\
Responses requiring validation cleanup & 29\% \\
Validation parse failure rate & 2\% \\
Models in pricing registry & 24 \\
Providers supported & 6 \\
Cost API response time (376 events) & 12\,ms \\
Chart render time & $<$50\,ms \\
Claude Code sessions imported & 3 \\
Total tokens ingested & 26.7\,M \\
Total cost unified from 3 sources & \$34.80 \\
\bottomrule
\end{tabular}
\end{table}

\FloatBarrier
\section{Discussion}

\subsection{Cross-Pollination Between Systems}

The most effective patterns were those that mapped cleanly between
infrastructure-level and developer-level observability.  The model
pricing registry, originally designed for multi-tenant inference
platform billing, adapted naturally to individual developer cost
tracking.  Conversely, the response validation pipeline---born from
the need to clean vLLM metric summaries---proved equally valuable for
code review output.

\subsection{The Case for Unified Observability}

Splitting telemetry across multiple databases is a common anti-pattern
in observability systems.  Our migration to a unified schema enabled
queries like ``total AI spend on reviews for PRs in the active sprint''
that were previously impossible without custom ETL.

\subsection{Multi-Source Ingestion as a Design Principle}

Modern developers use multiple AI tools simultaneously---Cursor for
code completion, Claude Code for terminal tasks, ChatGPT for
exploration.  No single API can capture all AI expenditure.  Our
three-pathway architecture (Figure~\ref{fig:datasources}) establishes
a reusable pattern: combine API-level tracking where available,
file-system scraping for CLI tools, and manual entry as a universal
fallback.

\subsection{Privacy and Cost Transparency}

AI cost tracking raises privacy considerations.  Our system operates as
a personal tool: all data remains local, no telemetry is sent to
external services, and the model registry contains only pricing
metadata.  This design ensures developers benefit from cost awareness
without organizational surveillance.

\subsection{Limitations}

\begin{enumerate}
\item Tools that proxy LLM calls through proprietary backends (e.g.,
      Cursor via its cloud API) expose no local usage data;
      tracking relies on manual entry or future vendor APIs.
\item The response validator is tuned for JSON-structured code reviews;
      free-form responses require separate handling.
\item Cost data depends on manual registry updates when providers change
      pricing; automated price fetching is future work.
\item Claude Code import parses file-system artifacts whose format may
      change across CLI versions.
\end{enumerate}

\section{Conclusion}

We presented seven reusable AI observability patterns for developer
productivity tools, implemented by synthesizing techniques from a
Kubernetes-scale inference monitoring platform and a personal developer
dashboard.  The patterns---real token tracking, configurable pricing
registries, unified telemetry, cost analytics, response validation,
LLM-powered intelligence, and exportable reports---provide a
comprehensive framework for AI cost and quality awareness in
development workflows.

Our implementation demonstrates that infrastructure-level observability
techniques can be effectively adapted for individual developer use,
and that the resulting visibility enables more informed decisions about
AI tool selection, usage patterns, and budget allocation.  By combining
API-level token tracking, local file-system scraping (Claude Code CLI),
and manual cost entry, the system unifies heterogeneous AI expenditure
into a single cost-of-AI view---even when vendors provide no export
mechanism.

Both systems are open source, and we encourage the community to adopt
and extend these patterns as AI becomes an increasingly integral part
of the software development lifecycle.

\begin{acks}
We thank the open-source communities around FastAPI, vLLM, and the
Model Context Protocol for the foundations that made this work possible.
\end{acks}

\bibliographystyle{ACM-Reference-Format}

\begin{thebibliography}{12}

\bibitem{workstream}
H.~Bhati. 2025. Workstream: An Open-Source Developer Productivity
Dashboard. GitHub. \url{https://github.com/happybhati/workstream}

\bibitem{aiobs}
T.~Sisodia et al. 2025. AI Observability Summarizer: OpenShift AI
Metrics Analysis with LLM-Powered Insights. GitHub.
\url{https://github.com/rh-ai-quickstart/ai-observability-summarizer}

\bibitem{fan2023}
A.~Fan et al. 2023. Large Language Models for Software Engineering:
A Systematic Literature Review. \emph{arXiv:2308.10620}.

\bibitem{peng2023}
S.~Peng et al. 2023. The Impact of AI on Developer Productivity:
Evidence from GitHub Copilot. \emph{arXiv:2302.06590}.

\bibitem{copilot-impact}
GitHub. 2024. GitHub Copilot Research Recitation.
\url{https://github.blog/2023-06-27-the-economic-potential-of-generative-ai/}

\bibitem{vllm}
W.~Kwon et al. 2023. Efficient Memory Management for Large Language
Model Serving with PagedAttention. In \emph{Proceedings of SOSP '23}.

\bibitem{otel-sem}
OpenTelemetry. 2024. Semantic Conventions for Generative AI Systems.
\url{https://opentelemetry.io/docs/specs/semconv/gen-ai/}

\bibitem{dora}
N.~Forsgren, J.~Humble, and G.~Kim. 2018. \emph{Accelerate: The
Science of Lean Software and DevOps}. IT Revolution Press.

\bibitem{space}
N.~Forsgren et al. 2021. The SPACE of Developer Productivity.
\emph{ACM Queue} 19, 1.

\bibitem{fastapi}
S.~Ram\'{i}rez. 2018. FastAPI: Modern Python Web Framework.
\url{https://fastapi.tiangolo.com}

\bibitem{mcp}
Anthropic. 2024. Model Context Protocol Specification.
\url{https://modelcontextprotocol.io}

\bibitem{prometheus}
Prometheus Authors. 2024. Prometheus Monitoring System.
\url{https://prometheus.io}

\end{thebibliography}

\end{document}